\documentclass[12pt]{article}
\usepackage{lscape} 
\usepackage{amsthm}
\usepackage{amsmath}
\usepackage{setspace}
\usepackage{csquotes} 
\usepackage{rotating}
\usepackage{fancyhdr}
\usepackage{graphicx}
\usepackage{ytableau}
\usepackage{url}
\usepackage{enumitem}
\usepackage{placeins}
\usepackage[labelfont=bf]{caption}
\usepackage{amssymb}
\usepackage{verbatim}
\usepackage{bbold}
\usepackage[export]{adjustbox}
\usepackage{cancel}
\usepackage{cite}
\usepackage{amsfonts}
\usepackage{tikz}
\usepackage{physics}
\usepackage[authoryear,longnamesfirst]{natbib}
\usepackage{color,xcolor,soul}
\usepackage[
    colorlinks=true,        
    linkcolor=black,       
    citecolor=black,       
    urlcolor=black,         
    bookmarks=true,         
    pdfauthor={Francesco Pio De Cosmo, Gianluca Di Natale, Leandro Cieri},
    pdftitle={ A Gauge-Inspired Coupled Fokker–Planck System for Multistate Stochastic Dynamics: Formalism and 1-D Numerical Implementation}
]{hyperref}

\begin{document}

\begin{titlepage}
\begin{flushright}
FTUV-26-1005.6123
\end{flushright}

\begin{center}
{\Large \bf  
  A Gauge-Inspired Coupled Fokker–Planck System \\for Multistate Stochastic Dynamics:\\ Formalism and 1-D Numerical Implementation \\
}
\end{center}

\par \vspace{2mm}
\begin{center}
  {\bf Francesco Pio De Cosmo${}^{(a,b)}$} {\bf Gianluca Di Natale${}^{(c)}$}, 
  and {\bf Leandro Cieri${}^{(d)}$}\\

\vspace{5mm}
${}^{(a)}$ 
University of Florence, Department of Information Engineering, Via di Santa Marta, 3, Firenze, Italy.\\\vspace{1mm}

${}^{(b)}$ 
Consiglio Nazionale delle Ricerche, Institute for Application of Calculous, Via Madonna del Piano, 10, Sesto Fiorentino, Firenze, Italy. \\\vspace{1mm}

${}^{(c)}$ 
Consiglio Nazionale delle Ricerche, National Institute of Optics, Via Madonna del Piano, 10, Sesto Fiorentino, Firenze, Italy.\\\vspace{1mm}

${}^{(d)}$ 
Instituto de F\'isica Corpuscular (IFIC), CSIC - Universitat de Val\`encia, Parc Cient\'ific, E-46980 Paterna, Valencia, Spain\\\vspace{1mm}
\end{center}

\vspace{1.5cm}

\par \vspace{2mm}
\begin{center}
{\large \bf Abstract}
\end{center} 
Extending the Fokker--Planck equation to a coupled system of equations through non-Abelian symmetry groups offers a powerful theoretical framework for coupled multi-state stochastic systems. However, translating this abstract gauge field formulation into viable numerical predictions remains an unaddressed challenge. Motivated by our previous work on ice crystal habit evolution, in this work we bridge this gap by presenting the first numerical solver for an $SU(2)$ gauge-coupled Fokker--Planck system in $(1+1)$ dimensions. Using an operator-splitting scheme combined with a non-oscillatory advection solver, our algorithm strictly ensures probability conservation, non-negativity, and numerical stability. We demonstrate that non-Abelian gauge coupling acts as a cross-reactive feedback loop that fundamentally alters system dynamics, driving continuous probability exchange between states, suppressing localised density peaks, and forcing long-time spatial delocalisation. Validated across linear and quadratic coupling regimes up to strong interactions ($g \lesssim 1.0$), this work establishes the foundational computational infrastructure needed to integrate complex multi-state physics. Overall, while our approach provides a key step toward improving ice cloud representation in atmospheric and climate models, the underlying physical framework and computational engine remain fundamentally general and multidisciplinary, offering a versatile paradigm for any domain requiring coupled multi-state stochastic equations. 

\begin{quote}
\pretolerance 10000

\end{quote}

\vspace*{\fill}
\vspace*{1.5cm}

\end{titlepage}

\setcounter{footnote}{1}
\renewcommand{\thefootnote}{\fnsymbol{footnote}}
\section{Introduction}\label{Intro}
The macroscopic time evolution of continuous stochastic processes in complex physical systems is universally governed by Fokker--Planck (FP) equations \cite{Risken1996}. This classical formulation is widely used across various disciplines, ranging from finance \cite{bouchaud2009theory} and neuroscience \cite{Brunel2000} to atmospheric physics and climate dynamics \cite{Hasselmann1976,Prusa2008,Majda2008}. However, modelling complex processes, such as the habit evolution of ice crystals in cirrus clouds, requires describing stochastic systems characterised by multiple correlated configurations. To address this complexity, advanced theoretical extensions are needed to transition from a single equation to a system of coupled equations. In our recent work \cite{DINATALE2026131312}, we established a novel theoretical framework addressing this challenge by mapping probability densities and physical currents into a gauge field.

Specifically, we demonstrate that ice crystal growth, traditionally categorised into canonical habits including droxtals, plates, columns, and bullet rosettes~\cite{UmMcFarquhar2007,Yang2015}, can be formally described by extending the classical Fokker--Planck equation, mappable to a $U(1)$ gauge theory following the approach of \cite{Montigny2003,Santana:2007Et}, through the elevation of its internal symmetry group, in analogy with the idea of synthetic non-Abelian gauge fields in optics~\cite{Chan2019}. While independent growth regimes are naturally captured by an Abelian $U(1)^N$ product symmetry,  where $N$ denotes the number of distinct habits, mutual morphological transitions and cross-talk between distinct crystal habits can be modelled by upgrading the framework to a non-Abelian $SU(2) \otimes U(1)$ gauge structure. This non-Abelian mapping induces a closed set of coupled Fokker--Planck equations containing explicit linear and quadratic interaction terms governed by a habit coupling constant $g_H$.

Although our foundational paper formulated the $SU(2) \otimes U(1)$ gauge-coupled Fokker--Planck formalism and outlined its physical significance for climate and radiative transfer modelling, explicit numerical solutions and algorithmic implementations of the coupled system had not yet been achieved. The primary objective of this paper is to bridge this critical gap, translating the abstract gauge-theoretic formalism into a robust, high-performance numerical engine. To achieve computational feasibility without sacrificing the fundamental non-linear mechanics of the theory, we restrict our investigation to a controlled $(1+1)$-dimensional (1 temporal + 1 spatial) proxy framework. This dimensional limitation serves as a rigorous benchmark to analyse how distinct probability sectors interact under non-Abelian symmetry constraints. The specific objectives of this contribution are threefold:

\begin{enumerate}
    \item Algorithmic validation: we develop and benchmark a customized non-oscillatory forward-in-time (NFT) operator-splitting scheme, validating its accuracy against exact analytical solutions of uncoupled Ornstein-Uhlenbeck processes \cite{PhysRev.36.823}. 
    
    \item Multistate non-linear dynamics: we numerically solve the $SU(2)$ gauge-coupled system of integro-differential equations, systematically exploring the role of the coupling strength $g$, initial condition asymmetries, and the asymptotic long-time behaviour that leads to global probability delocalization across states.
    
    \item Multidisciplinary applicability: We emphasise that while our investigation was originally motivated by cloud microphysics, the proposed framework based on coupled Fokker--Planck equations is formulated in full generality. Beyond our atmospheric application, it offers a highly versatile modelling paradigm with broad resonance across diverse scientific domains, ranging from statistical mechanics to financial mathematics and neuroscience, where coupled stochastic transitions and multi-state dynamics remain largely unexplored. By providing the first explicit numerical solutions for this gauge-inspired stochastic framework, this study lays the essential computational groundwork to investigate these complex coupled dynamics across different disciplines. Overall, it lays the foundation for integrating multi-habit microphysics into cloud modelling and Global Climate Models (GCMs), ultimately advancing the representation of cirrus cloud radiative forcing in climate predictions \cite{Baran2009,Intrieri2002}.

\end{enumerate}

\section{Theoretical framework}\label{Theory}

The macroscopic time evolution of a continuous stochastic process can be described by the Fokker--Planck equation. Let $\mathbf{x} = (x_1, \ldots, x_M)^T \in \mathbb{R}^M$ represent an $M$-dimensional state vector of stochastic variables. The time evolution of the probability density function $P(\mathbf{x}, t)$ is governed according to the general Fokker--Planck equation~\cite{Risken1996,Gardiner2009,Tome2015}:
\begin{equation}\label{FP_generic}
\frac{\partial P(\mathbf{x}, t)}{\partial t} = -\sum_{i=1}^{M} \partial_i \bigl(D_i(\mathbf{x})P(\mathbf{x}, t)\bigr) + \sum_{i,j=1}^{M} \mathbb{D}_{ij} \partial_i \partial_j P(\mathbf{x}, t) \,,
\end{equation}
where $\partial_i \equiv \partial / \partial x_i$. The drift vector $D_i(\mathbf{x})$ denotes the deterministic driving forces acting on the system, whereas the diffusion tensor $\mathbb{D}_{ij}$ parametrises the stochastic fluctuations. In line with standard formulations~\cite{VanKampen2007}, we assume the diffusion coefficients to be independent of the variables $\mathbf{x}$ and $t$.

The macroscopic framework established in Eq.~\eqref{FP_generic} is intimately connected to the underlying microscopic dynamics. Indeed, the Fokker--Planck equation can be directly derived from a set of coupled Langevin equations of the form:
\begin{equation}
\frac{\mathrm{d}x_i}{\mathrm{d}t} = D_i(x_1, x_2, \ldots, x_M) + \xi_i(t) \,,
\end{equation}
where the stochastic terms $\xi_i(t)$ represent Gaussian-Markovian (white) noise processes characterised by the following statistical moments:
\begin{equation}
\langle \xi_i(t) \rangle = 0 \,, \quad \langle \xi_i(t)\xi_j(t') \rangle = 2\mathbb{D}_{ij}\delta(t - t') \,,
\end{equation}
with $\delta(t-t')$ being the Dirac delta function and $\mathbb{D}_{ij}$ modulating the noise intensity.

As demonstrated in~\cite{Montigny2003}, the classical formulation of the Fokker--Planck equation can be embedded within a gauge field framework. Specifically, Eq.~\eqref{FP_generic} admits a formal mapping to an Abelian $U(1)$ gauge theory, wherein probability fields are associated with the gauge fields.\\
To describe complex transition dynamics in multi-state systems, such as ice crystal habit growth, we recently showed how the extension of this mapping to both the Abelian $U(1)^N$ and non-Abelian $SU(N)$ symmetries~\cite{DINATALE2026131312} allows us to capture inter-state dynamics. 
Through this generalization, the gauge covariant derivative naturally couples the system's probability sectors, providing a unified framework for inter-state cross-talk.

This mathematical structure yields a generalized, closed set of coupled Fokker--Planck equations. The resulting system models the joint probability distribution of several interacting states, providing a solid theoretical foundation for multistate stochastic dynamics, which serves as the formal basis for the description of the combined evolution of different crystal habits.

\subsection{Derivation of gauge-coupled Fokker--Planck equations}

We briefly review the results of Ref.~\cite{DINATALE2026131312} to demonstrate how the classical Fokker--Planck equation can be mapped into a classical field theory within an Abelian $U(1)$ gauge framework, and subsequently generalized. Let us define the coordinate derivative as $\partial_\mu = (\partial_t, \partial_i)$ and introduce the probability current density $J_i(\mathbf{x}, t)$. This allows us to recast the generic Fokker--Planck equation, Eq.~\eqref{FP_generic}, into a local continuity equation representing probability conservation:
\begin{equation}\label{FP_continuity}
\partial_t P(\mathbf{x}, t) = - \partial_i J_i(\mathbf{x}, t)  \,,
\end{equation}
where the physical current is defined as:
\begin{equation}\label{FP_current}
J_i(\mathbf{x}, t) = D_i(\mathbf{x})P(\mathbf{x}, t) - \mathbb{D}_{ij} \partial_j P(\mathbf{x}, t) \,.
\end{equation}
Within the gauge field theory context, we can map this equation by starting from a pure gauge theory framework, structurally analogous to classical electromagnetism. By implementing an appropriate choice of gauge-fixing condition, the corresponding field equations of motion simplify directly to the compact form:
\begin{equation}\label{gauge_fixing}
\partial_\mu A^\mu = 0 \,.
\end{equation}
Under this mathematical formulation, we identify the components of the gauge vector potential $A_\mu = (A_0, A_i)$, hereafter referred to as the Fokker--Planck (FP) field, by means of the following parametrization in terms of the probability density and its associated physical current:
\begin{equation}\label{Abelian_mapping}
A_0(\mathbf{x}, t) = P(\mathbf{x}, t) \,, \quad A_i(\mathbf{x}, t) = J_i(\mathbf{x}, t) \,.
\end{equation}
This mapping can be generalized to treat multi-configuration systems by upgrading the symmetry group to a non-Abelian $SU(N)$ gauge structure. By generalising this approach to a non-Abelian $SU(N)$ symmetry framework, one can implement an analogous gauge-fixing condition, whereby solving the corresponding field equations of motion directly yields the interacting system for the components $P_a$ (where $a=1,2,\ldots,N^2-1$ span the algebra generators): 
\begin{equation}\label{FPabelian}
\frac{\partial P_a}{\partial t}+ \partial_i\bigl(D_{(a)}^{i}P_a\bigr) - \mathbb{D}_{(a)}^{ij}\partial_i \partial_j P_a = f_{abc} \bigl(g\,\mathcal{I}^{(1)}(P^b,P^c) + g^2\,\mathcal{I}^{(2)}(P^b, P^c)\bigr) \,,
\end{equation}
where $f_{abc}$ denote the structure constants of the group, $g$ represents the coupling strength parametrization, and the functional interaction terms $\mathcal{I}^{(1)}$ and $\mathcal{I}^{(2)}$ are defined as \citep{DINATALE2026131312}\footnote{We point out a typographical error in our previous work \cite{DINATALE2026131312}, where the first term in parentheses in the linear coupling integral $\mathcal{I}^{(1)}$ was introduced with an incorrect minus sign.}:
\begin{equation}\label{1D_coupled_general}
\begin{aligned}
    \mathcal{I}_{bc}^{(1)} (\mathbf{x},t) & = \int {\rm d}x^\nu  \left[  A_\mu^c\,\left(  \partial^\mu A^\nu_b - \partial^\nu A^\mu_b\right) - \partial_\mu\left( A^\mu_b A^\nu_c\right)\right] \,,\\
    \mathcal{I}_{bc}^{(2)} (\mathbf{x},t) & = - \int {\rm d}x^\nu  f_{blm}\, A^\mu_m \, A^\nu_l \, A_\mu^c \,.
\end{aligned}
\end{equation}
By readjusting the structure indices, we can reformulate the first term as follows:
\begin{equation}
\begin{aligned}
    \mathcal{I}^{(1)}_{bc} (\mathbf{x},t) & = 
     \int {\rm d}x^\nu     \left[ - A^c_\mu(\partial^\nu A^\mu_b)  - (\partial_\mu A^\mu_b) A^\nu_c  + 2 A_\mu^c \ (\partial^\mu A^\nu_b)\right]\,\\
\end{aligned}
\end{equation}
Consistently with the Abelian case, the non-Abelian gauge fields $A^\mu_a$ are directly mapped to the multi-state probability densities and their associated spatial currents via:
\begin{equation}\label{fieldA_FP}
\begin{aligned}
    A^i_a(\mathbf{x},t) &=  D^{{i}}_{(a)}(\mathbf{x}) P_a(\mathbf{x},t)  -  \mathbb{D}_{(a)}^{ij}\partial_j   P_a(\mathbf{x},t) \equiv J^i_a(\mathbf{x},t)\,\,,\\
    A^0_a(\mathbf{x},t) &= P_a(\mathbf{x},t)\,\,.      
\end{aligned}    
\end{equation}
Explicitly expanding these integrals in terms of the physical currents yields:
\begin{equation}\label{currentsI}
\begin{aligned}
 \mathcal{I}^{(1)}_{bc} (\mathbf{x},t) &=   \int {\rm d}x^i \left[ - J^c_j(\partial^i J^j_b)  - (\partial_j J^j_b) J^i_c + 2 J_j^c \ (\partial^j J^i_b) - P^c(\partial^i P_b)  - (\partial_t P_b) J^i_c + 2 P^c \ (\partial_t J^i_b)\right] \\
 &+  \int {\rm d}t \left[ - J^c_i(\partial_t J^i_b)  - (\partial_i J^i_b) P_c + 2 J_i^c \ (\partial^i P_b)\right] \,, \\
 \mathcal{I}^{(2)}_{bc} (\mathbf{x},t) &= - \int {\rm d}x^i f_{blm}\,(  P_m \, J^i_l \, P^c +  J^\mu_m \, J^i_l \, J_\mu^c  ) - \int {\rm d}t f_{blm}\,(  P_m \, P_l \, P^c +  J^\mu_m \, P_l \, J_\mu^c  ) 
\end{aligned}
\end{equation}
For more details, see the Appendix \ref{Appendix_Integrals}.
\subsection{The (1+1)-dimensional $SU(2)$ framework}

The generalised system of coupled non-linear equations presented in Eq.~\eqref{FPabelian} exhibits a high degree of analytical complexity. This is fundamentally due to the non-local nature of the integrals $\mathcal{I}^{(1)}$ and $\mathcal{I}^{(2)}$, combined with the intrinsic non-Abelian self-interactions mediated by the structure constants $f_{abc}$ of the $SU(N)$ Lie algebra. In the full $M$-dimensional spatial formulation, the dynamics are governed by non-linear partial integro-differential equations where spatial gradients, drift vectors, diffusion tensors, and time-dependent integral constraints are deeply interconnected. Consequently, an exact analytical solution cannot be found, and thus necessitates a numerical approach.

With the ultimate goal of implementing this theoretical framework within our target physical context, the evolution of ice crystal habits in cirrus clouds under a pure vapor deposition regime, we restrict the theory to a controlled, simplified scenario. Projecting the theory into a $(1+1)$-dimensional space provides a fundamental proxy. This dimensional reduction serves as a necessary, computationally tractable step designed to benchmark how distinct probability sectors interact under the non-Abelian symmetry constraint. Once the one-dimensional coupling mechanics are firmly established and numerically verified, the framework is mathematically structured to be generalized to higher spatial coordinates ($D > 1$) and to directly incorporate the specific physical thermodynamic variables that drive the actual growth front.

To formally execute this dimensional reduction, we assume that the spatial-type dynamics are constrained along a single coordinate axis, denoted by $x$. Consequently, the system drops its spatial-type index dependencies, meaning that all spatial tensors and vectors collapse to single scalar components ($i,j \rightarrow x$). The differential operators scale accordingly as $\partial_i \rightarrow \partial_x$ and $\partial_i \partial_j \rightarrow \partial_x^2$. In this condition, the generalized multi-state physical currents $J^i_a(\mathbf{x}, t)$ lose their vector nature and reduce to the scalar spatial currents:
\begin{equation}\label{1D_current}
J_a(x, t) = D^x_{(a)}(x) P_a(x, t) - \mathbb{D}^{xx}_{(a)} \partial_x P_a(x, t) \,,
\end{equation}
where $D^x_{(a)}(x)$ and $\mathbb{D}^{xx}_{(a)} $ represent the one-dimensional drift term and the scalar diffusion coefficient for the $a$-probability sector, respectively. 

This reduction directly impacts the structural mapping of the gauge fields. The components of the non-Abelian vector potential $A^\mu_a = (A^0_a, A^i_a)$ contract to a simple two-component field structure $A^\mu_a = (P_a, J_a)$. By explicitly substituting these scalar field identifications into the functional interaction integrals $\mathcal{I}^{(1)}$ and $\mathcal{I}^{(2)}$, the spacetime volume element contracts to $\mathrm{d}x^\nu \rightarrow (\mathrm{d}t, \mathrm{d}x)$. 
Now we focus on the $SU(2)$ gauge group, which provides the algebraic foundation for our model. The $SU(2)$ Lie algebra is characterized by three generators ($a,b,c \in \{1,2,3\}$), allowing the structure constants to map directly into the totally anti-symmetric Levi-Civita tensor, namely $f_{abc} = \epsilon_{abc}$. 
As a result of this systematic reduction, the explicit equation for each coupled probability density component $P_a$ by using Eqs. (\ref{fieldA_FP}) and (\ref{currentsI}) takes the following form:
\begin{equation}\label{eq:1D_coupled_general}
\begin{aligned}
\partial_t P_a &= -\partial_x J_a \\
&+ g \ \epsilon_{abc} \left[ \int_{\Omega_x} \mathrm{d}x' \, \big( -P_c \partial_{x'} P_b  - (\partial_t P_b) J_c +  2 P_c (\partial_t J_b)\big) + \int_0^t \mathrm{d}t' \, \big( -J_c \partial_{t'} J_b - (\partial_x J_b) P_c + 2 J_c (\partial_x P_b) \big) \right] \\
& + g^2 \left[ \int_{\Omega_x} \mathrm{d}x' \, P_b ( P_b J_a -P_a J_b ) + \int_0^t \mathrm{d}t' \, J_b (  J_b P_a -J_aP_b) \right] \,.
\end{aligned}
\end{equation}
To make the underlying coupling dynamics transparent, we expand the field equations for each individual index component. 
Within the $SU(2)$ algebra, due to the properties of the Levi-Civita symbol, fixing the index $a$ restricts the indices $b$ and $c$ to range exclusively over the remaining two available values. For instance, choosing the first probability sector ($a=1$) constrains the active configurations to $b,c \in \{2,3\}$. By executing the explicit permutations over these active index pairs and tracking the sign changes induced by the Levi-Civita tensor, the algebraic summations expand into the following highly structured, coupled system of non-linear integro-differential equations:
\begin{equation}\label{SU2_system_explicit}
\begin{cases}
\begin{aligned}
\partial_t P_1 &= -\partial_x J_1 \\
&+ g \left[ \int_{\Omega_x} \mathrm{d}x' \, \big( -P_3 \partial_{x'} P_2  - (\partial_t P_2) J_3 +  2 P_3 (\partial_t J_2)\big) + \int_0^t \mathrm{d}t' \, \big( -J_3 \partial_{t'} J_2 - (\partial_x J_2) P_3 + 2 J_3 (\partial_x P_2) \big) \right] \\
& - \Big( 2 \leftrightarrow 3 \Big) + g^2 \left[ \int_{\Omega_x} \mathrm{d}x' \, P_2 \big( P_2 J_1 - P_1 J_2  \big) + \int_0^t \mathrm{d}t' \, J_2 \big(  J_2 P_1 - J_1 P_2\big) \right] + \Big( 2 \leftrightarrow 3 \Big) \,,
\end{aligned} \\
\quad \dots  
\end{cases}
\end{equation}
where $ (2 \leftrightarrow 3) $ denotes the identical integral blocks evaluated under index permutation. \\
This explicit reduction demonstrates that under an $SU(2)$ gauge symmetry, the non-linear coupling behaves as a conservative, cross-reactive feedback loop between the distinct states, providing a baseline for the numerical implementations we will discuss in the next section.

\section{Numerical methods and algorithmic implementation}

To solve the Fokker--Planck equation (FPE) numerically, various computational techniques can be found in the literature \cite{Risken1996,VanKampen2007}, each focus to more specific physical and mathematical problems. These methodologies uses Monte Carlo simulations \cite{Varje2015, Kikuchi1991}, the Finite Element Method (FEM) \cite{Kumar2006}, and robust finite-difference formulations for stochastic acceleration phenomena \cite{Park1996}. Other approaches rely on operator splitting strategies designed for low-dimensional dynamic environments \cite{Schmidt2009}, or more advanced frameworks utilizing Proximal Recursion \cite{Caluya2018}.

In this case, the target is the numerical solution of the non-linear, gauge-inspired coupled Fokker--Planck system established in Section~\ref{Theory}. As outlined previously, this theoretical framework is designed to model multi-state stochastic dynamics, specifically targeting the microphysical evolution of ice crystal habits under vapor deposition regimes in cirrus clouds \cite{Solch2010}. To solve this complex system of integro-differential equations and evaluate the cross-talk between distinct probability sectors, we implement a customized non-oscillatory forward-in-time (NFT) operator-splitting scheme within a reduced-order (1+1)-dimensional framework. Specifically, we first establish a benchmark baseline by validating this scheme on an uncoupled Ornstein-Uhlenbeck process and then we extend its algorithmic core to handle the non-Abelian gauge interactions. 

To achieve this, we introduce a numerical scheme directly inspired by the work of Kärcher \cite{Karcher2003,Karcher2014} within the context of ice cloud microphysics, using numerical solvers originally developed for geophysical fluids and multiscale atmospheric flow applications \cite{Prusa2008}. This methodology divides the continuous evolution of the Fokker--Planck equation into two distinct mathematical components referred to the drift and the diffusion terms. Specifically, the drift one is solved via the MPDATA algorithm \cite{Smolarkiewicz1998, Smolarkiewicz2006}, while the  diffusion term is calculated using a standard second-order central finite difference scheme. As discussed by Kärcher \cite{Karcher2003}, the MPDATA method offers three key structural advantages for habit modelling in this type of system: it preserves the strict non-negativity of the probability distribution, ensures the exact conservation of total probability without requiring arbitrary corrections, and eliminates non-physical numerical oscillations.

\subsection{Stability tests of the numerical code for the (1+1)-dimensional Fokker--Planck equation}
As a validation benchmark, we focus on a damped harmonic oscillator, formally described by an Ornstein-Uhlenbeck process, for more details see \cite{Risken1996}. In this configuration, the standard Fokker--Planck equation assume the form:
\begin{equation}\label{standard_FPE}
\partial_t P(x,t) = \partial_x \left[ \gamma x P(x,t) \right] + D  \partial_x^2 P(x,t) \,,
\end{equation}
where $D^x(x) = -\gamma x$ represents the linear spatial drift velocity, $\mathbb{D}^{xx}=D$ defines the constant diffusion coefficient, and in this context, $\gamma$ and $D$ are parameters that will be fixed during the simulation.  

In standard configurations, standard finite-difference approaches such as the Forward-Time Central-Space (FTCS) scheme are often deployed. However, when the drift term is characterized by sharp gradients or high advection regimes, the FTCS central-space discretization introduces non-physical numerical oscillations and severe artificial diffusion. Furthermore, explicit formulations are bound to strict Courant-Friedrichs-Lewy (CFL) stability conditions ($\Delta t \le \Delta t_{\max} = \Delta x^2 / (2 \mathbb{D}^{xx})$) \cite{strikwerda2004} that couple the temporal and spatial step lengths, and do not natively preserve the total probability integral, requiring cumbersome numerical normalization steps over time \cite{milstein2004}.

We thus implement an explicit, splitting method following Kärcher \cite{Karcher2003}, which decouples the drift and diffusion operators. For the numerical setup, the probability density function, $P(x, t)$, is discretized in time $t$ and in the $x$ space with finite differences: $P^n_i$ denotes the discrete dependent function, the upper index indicates the discretized time, $t^n = n\Delta t$, and the lower index the equally spaced grid points, $x_i = i\Delta x$ (with $i = 1, \dots, N_x$). All numerical simulations are performed on a finite computational domain \(\Omega_x=[x_{\min},x_{\max}]\), discretized using \(N_x\) uniformly spaced grid points. Spatial integrals and probability normalization are evaluated over this interval. Accordingly, the broadening and delocalization discussed below refer to redistribution within the finite computational domain.

Under this scheme, the integration over a time interval $t^n \to t^{n+1}$ factors the equations into a two-step sequence. In the first step, following the MPDATA algorithm approach the physical diffusion operator is computed to define an intermediate probability density $\widetilde{P}$ evaluated from the current time level $t^n$:
\begin{equation}\label{diffusion_precondition}
\widetilde{P}^n_i = P^n_i + \Delta t \, R^n_i \,.
\end{equation}
Here, $R^n_i$ represents the second-order central difference approximation used to discretize the spatial diffusion term:
\begin{equation}\label{diffusion_discretization}
R^n_i = \mathbb{D}^{xx} \frac{P^n_{i+1} - 2P^n_i + P^n_{i-1}}{\Delta x^2} \,.
\end{equation}
Boundary conditions for this intermediate profile are set to enforce local balance ($\widetilde{P}_0 = \widetilde{P}_1$, $\widetilde{P}_{N-1} = \widetilde{P}_{N-2}$). In the second step, this regularized intermediate state is transported and updated via the MPDATA drift scheme to obtain the final profile at the next time level $t^{n+1}$:\begin{equation}\label{NFT_operator_sequence}P^{n+1}_i = \mathcal{A}_i \left( t^{n+1/2}, \widetilde{P}^n \right) \,,\end{equation}where $\mathcal{A}$ represents the non-oscillatory MPDATA drift transport operator. Finally, the total probability distribution is normalized to ensure that its total integral remains strictly conserved.

The performance and mathematical consistency of the numerical methods are evaluated through two complementary validation steps.

\begin{figure}[t]
    \centering
    \includegraphics[width=0.95\textwidth]{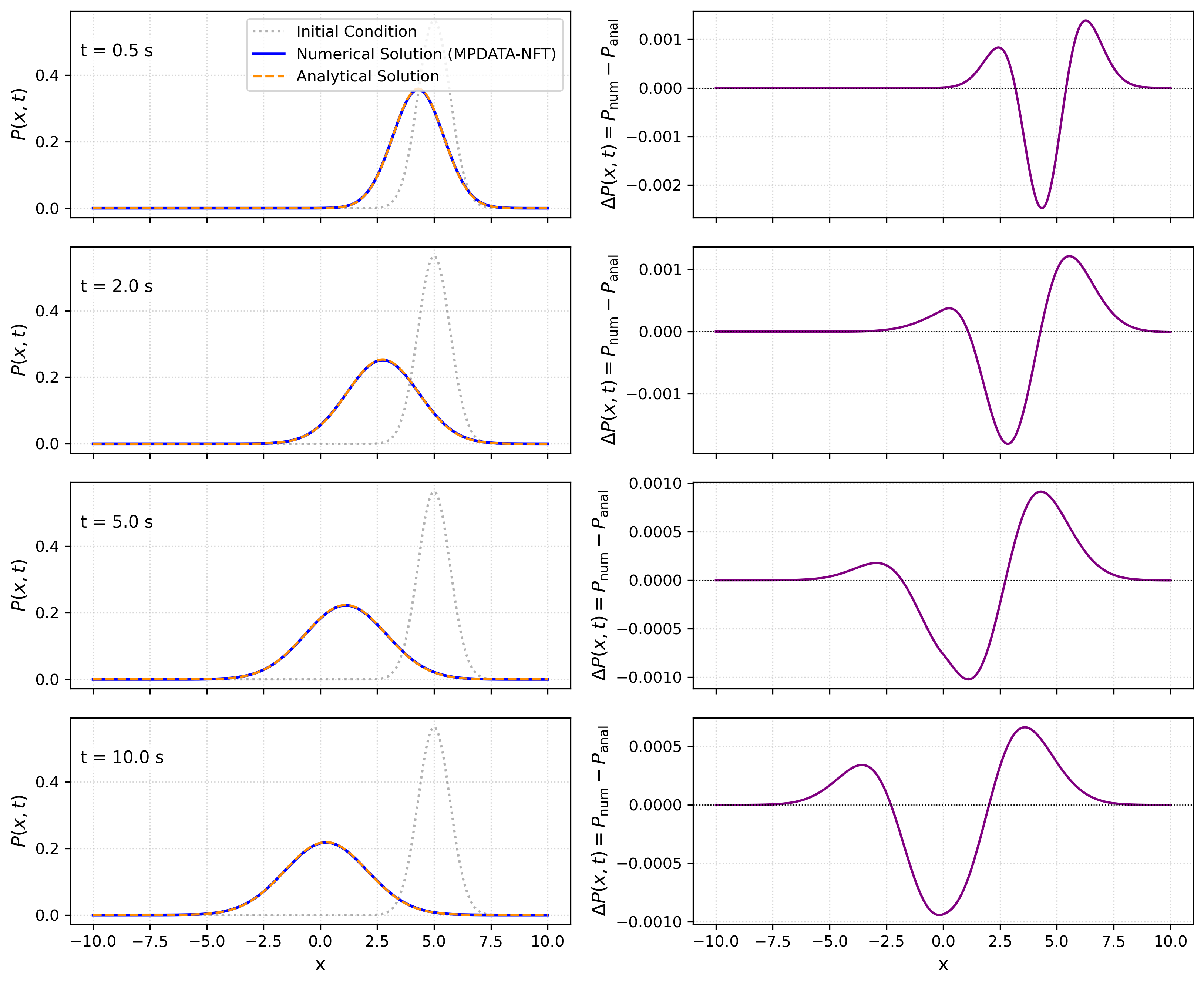}
    \caption{Time evolution of the probability density function $P(x,t)$ (left panels) and the corresponding spatial profile of the local numerical error $\Delta P(x,t) = P_{\mathrm{num}} - P_{\mathrm{anal}}$ (right panels) sampled at four distinct tracking intervals: $t = \{0.5, 2.0, 5.0, 10.0\}$ s. The dashed lines denote the exact analytical solution of the Ornstein-Uhlenbeck process, establishing the non-oscillatory and shape-preserving properties of the MPDATA-NFT solver.}
    \label{fig:profiles_and_errors}
\end{figure}

First, we compare the numerical solution against the exact analytical solution to evaluate the absolute accuracy of the split-operator sequence. Since, as mentioned, we consider an Ornstein-Uhlenbeck process, starting from a localized Gaussian initial profile centered at $\mu_0 = 5.0$ with a variance $\sigma^2_0 = 0.5$:
\begin{equation}\label{OU_initial_condition}
P(x, 0) =  \frac{1}{\sqrt{2\pi\sigma^2_0}} \exp\left( -\frac{(x - \mu_0)^2}{2\sigma^2_0} \right) \equiv \mathcal{N}(\mu_0, \sigma_0^2) \,,
\end{equation}
under a linear drift $D^x(x) = -\gamma x$ and constant diffusion $D$, the exact time-dependent analytical solution, used to evaluate the numerical error, is given by:
\begin{equation}\label{OU_analytical_solution}
P_{\mathrm{analytical}}(x, t) = \frac{1}{\sqrt{2\pi\sigma^2(t)}} \exp\left( -\frac{(x - \mu(t))^2}{2\sigma^2(t)} \right) \,,
\end{equation}
where the time-evolving mean $\mu(t)$ and variance $\sigma^2(t)$ are defined as:
\begin{equation}\label{OU_moments}
\mu(t) = \mu_0 e^{-\gamma t} \,, \quad \sigma^2(t) = \frac{D}{\gamma} \left( 1 - e^{-2\gamma t} \right) + \sigma^2_0 e^{-2\gamma t} \,.
\end{equation}
The numerical simulation is performed by fixing the physical benchmark parameters to $\gamma = 0.3$ and $D = 1.0$. As shown in Fig.~\ref{fig:profiles_and_errors}, the numerical solution successfully tracks the evolution of the distribution as it progresses toward the stationary state centered at $x=0$. The matching between the numerical solution (solid blue line) and the exact analytical solution (dashed orange line) is highly accurate across all sampled time intervals. 

The spatial structures of the local error, shown in the right-hand panels, exhibit distinctive lobe-like shapes that directly mirror the spatial derivatives of the Gaussian distribution. During the evolution, these lobes emerge as a mathematical signature of a shifting error, where the numerical solution is slightly misaligned with the analytical one. Quantitatively, the maximum absolute error is bounded near $2.5 \times 10^{-3}$ at early stages ($t = 0.5$ s) and smoothly decays below $9.6 \times 10^{-4}$ as the system relaxes toward equilibrium ($t = 10.0$ s), keeping the peak relative error under $0.72\%$. Furthermore, the profiles approach zero smoothly at the boundaries without generating non-physical negative values, confirming the overall robustness and numerical stability.

\begin{figure}[htbp]
    \centering
    \includegraphics[width=0.65\textwidth]{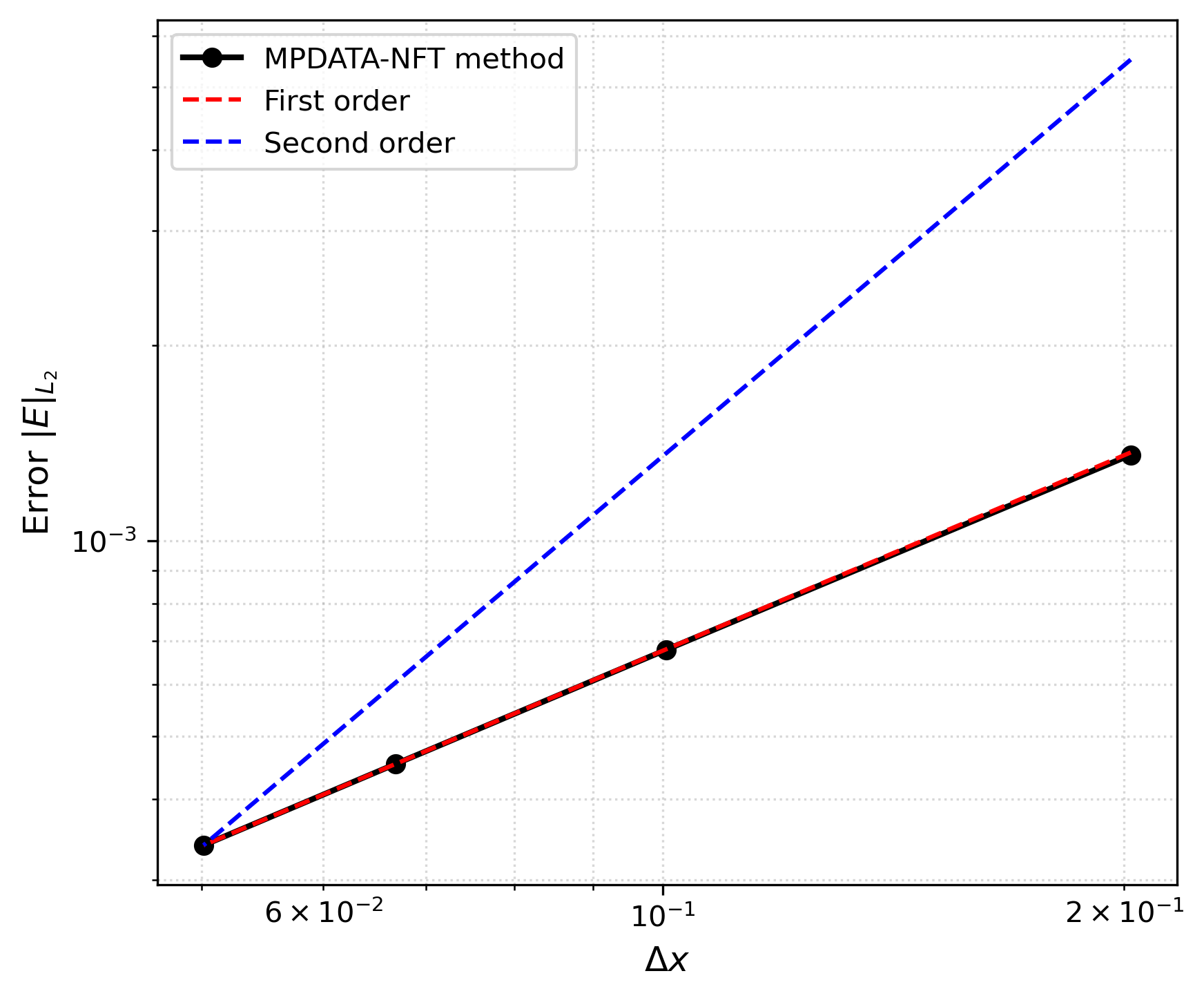}
    \caption{Log-log grid convergence study evaluating the global $L_2$ error norm at $t = t_{\mathrm{final}}$ as a function of the spatial grid spacing $\Delta x$. The empirical computational refinement (solid black line) is benchmarked against asymptotic first-order ($\mathcal{O}(\Delta x^1)$, dashed red line) and second-order ($\mathcal{O}(\Delta x^2)$, dashed blue line) slopes, illustrating how spatial accuracy degrades to first-order due to the numerical dissipation inherent in un-centered explicit operator-splitting schemes.}
    \label{fig:grid_convergence}
\end{figure}
We next perform a more systematic analysis for four independent grid topologies $N_x = \{100, 200, 300, 400\}$, as illustrated in Fig.~\ref{fig:grid_convergence}. A log-log linear regression mapping the global $L_2$ error norm at the final time against the spatial grid spacing $\Delta x$ yields an empirical computational slope of $1.02$. 

The $L_2$ error norm quantifies the discrepancy between the numerical solution $P_{\mathrm{num}}(x, t_{\mathrm{final}})$ and the exact analytical solution $P_{\mathrm{anal}}(x, t_{\mathrm{final}})$ across the entire spatial domain $\Omega$. Discretized over the grid nodes $i = 1, \ldots, N_x$, it is formally defined as: 
\begin{equation}\label{L2_error_definition}
\| \mathcal{E} \|_{L_2} = \sqrt{ \sum_{i=1}^{N_x} \left| P_{\mathrm{num}}(x_i, t_{\mathrm{final}}) - P_{\mathrm{analytical}}(x_i, t_{\mathrm{final}}) \right|^2 \Delta x } \,,
\end{equation}
where the multiplication by $\Delta x$ acts as the discrete integration volume element, ensuring that the error norm remains consistent under grid refinement. 

Although the standalone MPDATA scheme exhibits formal second-order spatial accuracy \cite{Smolarkiewicz1998, Karcher2003},  decoupling the drift and diffusion terms within an explicit fractional-step splitting scheme restricts the overall framework to a first-order accuracy regime. This behaviour is confirmed by the error curve (solid black line), which perfectly aligns  with the first-order reference slope (dashed red line) and clearly departs from the second-order case (dashed blue line). This match between the measured empirical slope and the theoretical expectation successfully validates the mathematical consistency and spatial accuracy of the implemented numerical scheme.

\subsection{Gauge-coupled Fokker--Planck system}

Following the validation of the numerical scheme under the uncoupled Fokker--Planck benchmark, we implement the system of equations defined in Eq.~\ref{SU2_system_explicit}, passing from a single equation to a system of $N = 3$ coupled integro-differential equations. The underlying Lie algebra introduces a fundamental mathematical challenge already formalized in the general $1$D reduction of Eq. (\ref{1D_coupled_general}): the probability fields are coupled through cyclic, non-local source terms where the evolution of each specific distribution depends on the properties of the other sectors.

To account for the additional dynamical terms arising from the non-Abelian $SU(2)$ extension, the time evolution of the system is now governed also by non-local memory effects. Mathematically, evaluating these non-local integrals requires a proper discretization of the fields and their derivatives. For this purpose, we define the local physical current $J_{a, i}^n$ as:
\begin{equation}\label{flux_discretization}
J_{a, i}^n = D^x_{(a)}(x_i)P_{a, i}^n - \mathbb{D}_{(a)}^{xx} \left[ \frac{P_{a, i+1}^n - P_{a, i-1}^n}{2\Delta x} \right] \,, 
\end{equation}
while the temporal derivatives are approximated via a backward finite difference scheme:
\begin{equation}\label{temporal_derivatives}
\partial_t P_{a, i}^n \approx \frac{P_{a, i}^n - P_{a, i}^{n-1}}{\Delta t} \,, \qquad \partial_t J_{a, i}^n \approx \frac{J_{a, i}^n - J_{a, i}^{n-1}}{\Delta t} \,.
\end{equation}
The equations now incorporate two distinct classes of non-local coupling terms. 
The first category comprises spatial interaction terms that represent domain-wide constraints. While spatially uniform across the domain $\Omega$, these terms evolve dynamically over time and are computed using a discrete trapezoidal integration rule following Eq. (\ref{eq:1D_coupled_general}):
\begin{align}
\Gamma_{a, \text{space}, \, g}^n &= \sum_{b,c=1}^{3} \epsilon_{abc} \int_{\Omega} \left[ P_b \partial_x P_c - 2 P_b \partial_t J_c + \partial_t P_c J_b \right] \mathrm{d}x \,, \label{num_gamma_space_g} \\
\Gamma_{a, \text{space}, \, g^2}^n &= \sum_{\substack{b=1 ,\\ b \neq a}}^{3}  \int_{\Omega} \left[ P_b \left( P_b J_a -P_a J_b \right) \right] \mathrm{d}x \,. \label{num_gamma_space_g2}
\end{align}

Simultaneously, the second class involves non-local temporal memory sources that act as localized variables tracking the evolution of the system. Our strategy to evaluate these memory terms consists of storing the integrated timelines and updating them recursively at each discrete time level:
\begin{align}
\Theta_{a, i, \, g}^{\text{time}}(t_n) &= \Theta_{a, i, \, g}^{\text{time}}(t_{n-1}) + \Delta t \sum_{b,c=1}^{3} \epsilon_{abc} \left[ J_{b, i} \partial_t J_{c, i} + \partial_x J_{c, i} P_{b, i} - 2 J_{b, i} \partial_x P_{c, i} \right] \,, \label{num_theta_time_g} \\
\Theta_{a, i, \, g^2}^{\text{time}}(t_n) &= \Theta_{a, i, \, g^2}^{\text{time}}(t_{n-1}) + \Delta t \sum_{\substack{b=1 ,\\ b \neq a}}^{3}  \left[ J_{b, i} \left( J_{b, i} P_{a, i} - J_{a, i} P_{b, i} \right) \right] \,, \label{num_theta_time_g2}
\end{align}
where the spatial derivatives of the currents are evaluated via central differences, the fields $J$ and $P$ the temporal index $n$ for brevity, and memory terms are initialized to zero at $t_0$. 

The diffusion term $R^n_{a,i}$ is discretized across the internal grid nodes $i = 1, \ldots, N_x-2$, following the central difference approximation used in the previous section, as:
\begin{equation}\label{SU2_diffusion_discretization}
R^n_{a,i} = \mathbb{D}_{(a)}^{xx} \frac{P^n_{a, i+1} - 2P^n_{a, i} + P^n_{a, i-1}}{\Delta x^2} \,.
\end{equation}the splitting, these new coupling source terms and the second-order diffusion term are evaluated first. This combined source-diffusion step updates the probability density fields from the previous time level $t^n$ to generate a set of intermediate, pre-conditioned profiles $\widetilde{P}_{a,i}$:
\begin{equation}\label{SU2_precondition}
\widetilde{P}^n_{a,i} = P^n_{a,i} + \Delta t \left[ R^n_{a,i} + g \cdot \Gamma_{a, \text{space}, \, g}^n + g \cdot \Theta_{a, i, \, g}^{\text{time}}(t_n) + g^2 \cdot \Gamma_{a, \text{space}, \, g^2}^n + g^2 \cdot \Theta_{a, i, \, g^2}^{\text{time}}(t_n) \right] \,.
\end{equation}
Boundary conditions for these intermediate profiles are immediately set to enforce local balance before executing the transport step ($\widetilde{P}_{a, 0} = \widetilde{P}_{a, 1} $ and $\widetilde{P}_{a, N_x-1} = \widetilde{P}_{a, N_x-2}$), preventing unphysical boundary accumulation.

Once these intermediate states are defined, the pure advective transport induced by the spatially dependent drift fields $D^x_{(a)}(x)$ is resolved by applying the non-linear MPDATA transport operator. This final step advances the entire coupled system to the next temporal level $t^{n+1}$ while ensuring positive definiteness and preventing non-physical numerical oscillations. Under this operational sequence, the continuous evolution over a discrete time step $\Delta t$ factors the integration for each sector $a = 1, 2, 3$ as:
\begin{equation}\label{SU2_operator_sequence}
P^{n+1}_{a,i} = \mathcal{A}_i \left( t^{n+1/2}, \widetilde{P}^n_a \right) \,,
\end{equation}
where $\mathcal{A}$ represents the non-oscillatory MPDATA transport solver.

Finally, the non-linear coupling complexity requires a global renormalization step at each time level to ensure total probability conservation and non-negativity across all sectors. Specifically, each density is scaled by the total integrated probability of the multi-state domain to guarantee that the total sum is equal to $N$. This constraint extends the single-component conservation principle to the coupled gauge sectors, simultaneously balancing the minor localized discrepancies introduced by the splitting scheme and non-local source terms.

\section{Results}\label{Results}

In this section, we analyse the numerical simulations of the one-dimensional $SU(2)$ gauge-coupled Fokker--Planck system to evaluate the results of our numerical implementation, studying both the performance and the physical effects of this new coupling compared to the standard, uncoupled Fokker--Planck reference. To  explore the behaviour of the equations and the properties of their solutions, we focus our investigation on two key aspects. First, we evaluate the response of the system to different initial conditions, testing the dynamics of the equations under distinct starting configurations and comparing them directly to the uncoupled case. Second, we examine the role of the coupling strength $g$, analysing its effects as it scales from smaller to larger values, alongside the long-time limit condition, tracking how the interaction alters the asymptotic evolution of the system. We solve the system of equations~\eqref{eq:1D_coupled_general} considering Ornstein-Uhlenbeck-type processes with a linear drift $D_{(a)}^x(x) = -\gamma_a x$, and a constant diffusion coefficient $\mathbb{D}_{(a)}^{xx} = D_{a}$. The drift parameters $\{\gamma_a\}$ and diffusion coefficients $\{D_a \}$ are fixed during each simulation and for each of the three states. 

\subsection{Comparative analysis: uncoupled and coupled solutions}
\begin{figure}[b!]
    \centering
    % === IMMAGINE A SINISTRA (Case a) ===
    \begin{minipage}[t]{0.48\textwidth}
        \centering
        \includegraphics[width=\textwidth]{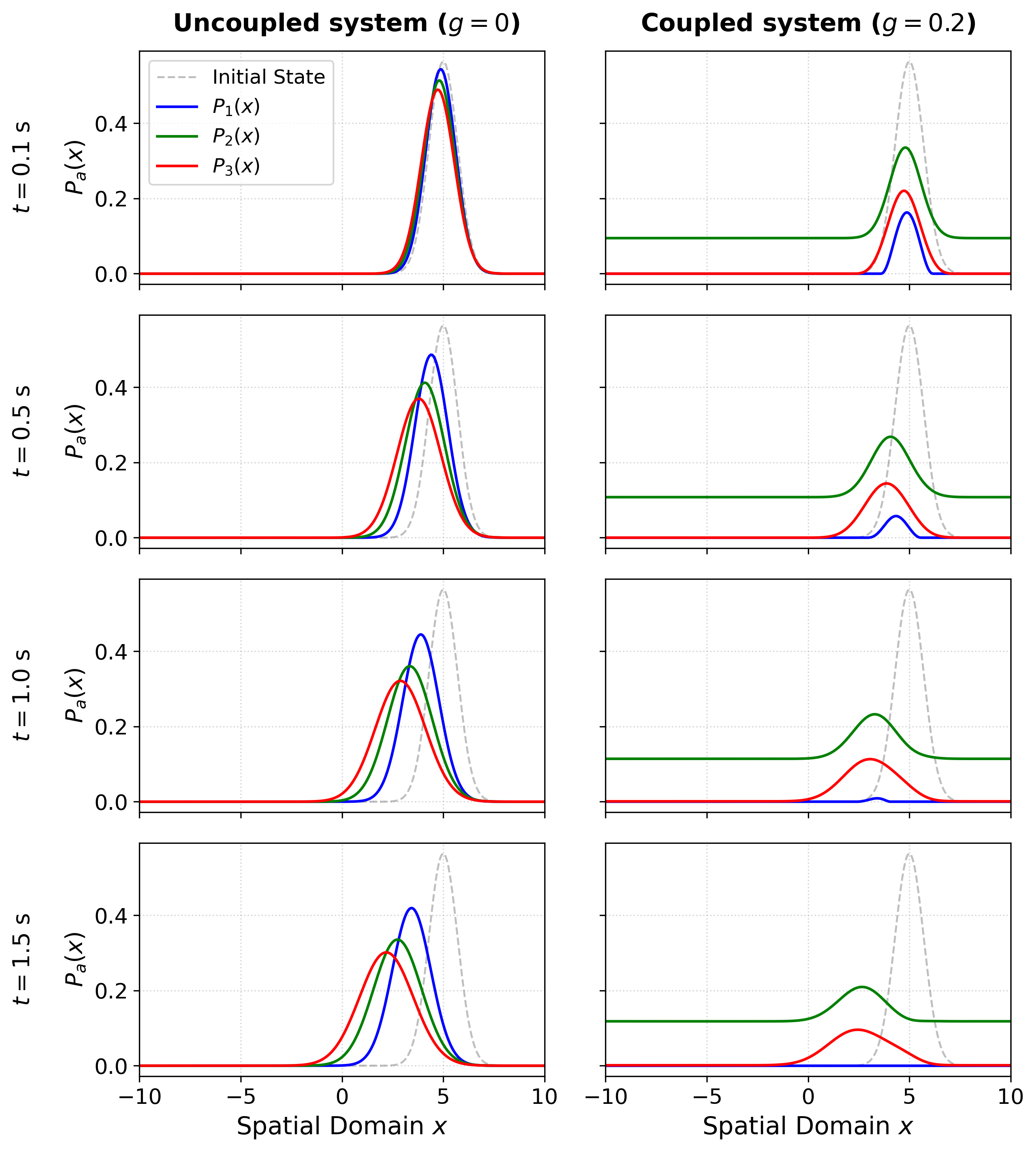}
        \vspace{1mm}
        \\ {(a)}
        \label{fig:case_a}
    \end{minipage}%
    \hfill 
    % === IMMAGINE A DESTRA (Case b) ===
    \begin{minipage}[t]{0.48\textwidth}
        \centering
        \includegraphics[width=\textwidth]{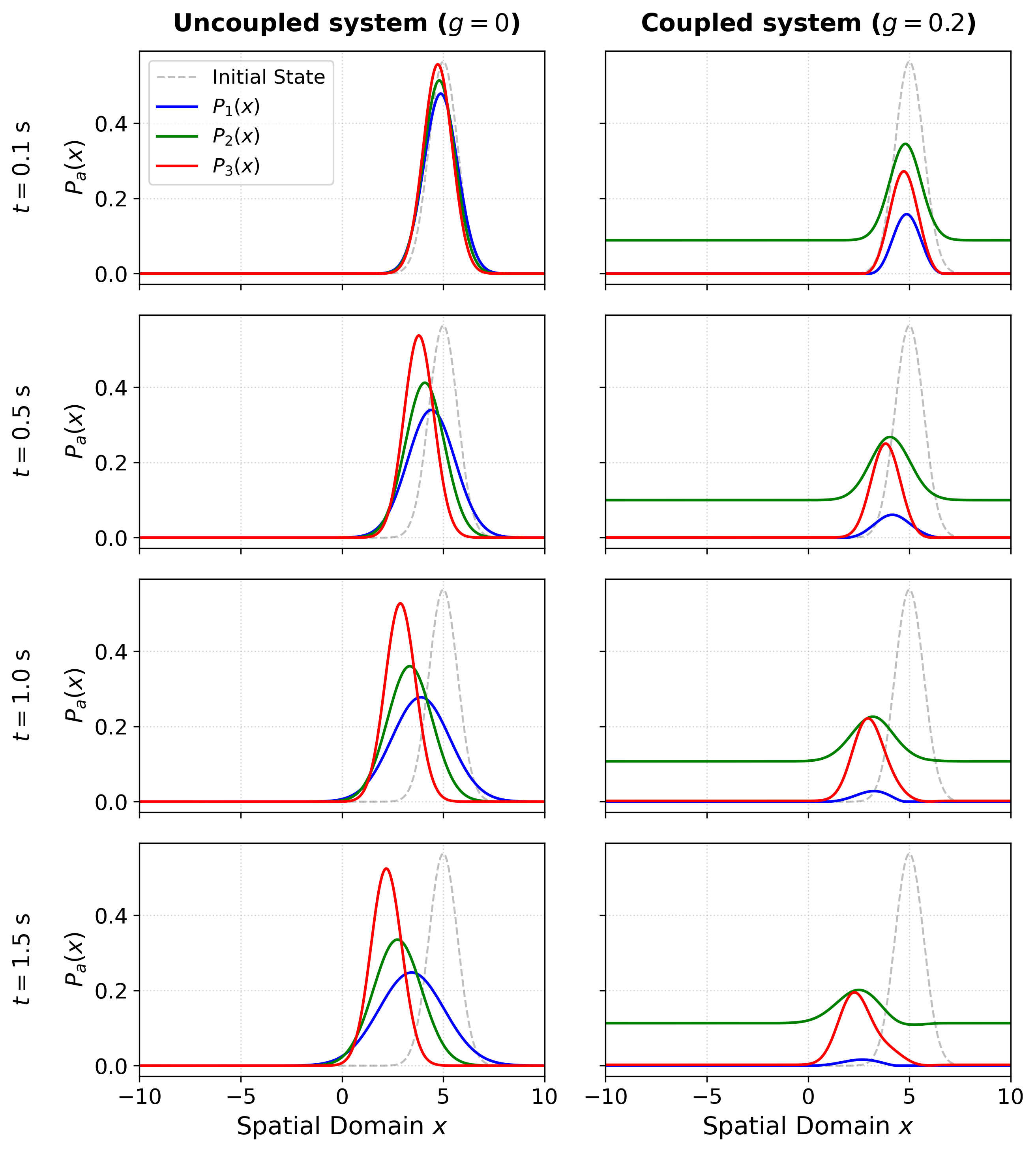}
        \vspace{1mm}
        \\ {(b)}
        \label{fig:case_b}
    \end{minipage}
    \caption{Temporal evolution of the probability density profiles $P_a(x)$ for the three sectors ($a=1,2,3$) (blue, green and red) starting from identical Gaussian initial profiles centered at $\mu_0 = 5.0$ with width $\sigma = 1.0$. The two panels represent two parameter configurations. The drift coefficients are ($\gamma_1 = -0.25x$, $\gamma_2 = -0.4x$, $\gamma_3 = -0.55x$). In Panel (a) the diffusion coefficients are $D_1 = 0.3$, $D_2 = 0.7$, and $D_3 = 1.1$, and in Panel (b) they are $D_1 = 1.1$, $D_2 = 0.7$, and $D_3 = 0.3$. In both panels, the left columns represent the uncoupled model ($g=0$) and the right columns illustrate the gauge-coupled model ($g=0.2$).}
    \label{fig:coupled_uncoupled_comparison}
\end{figure}

To evaluate a first qualitative effect of the proposed coupled equations, we compare the independent evolution of three decoupled Fokker--Planck equations against their complete gauge-coupled counterparts. In Fig.~\ref{fig:coupled_uncoupled_comparison}, we show the temporal evolution of the probability density profiles $P_a(x)$ for the three gauge indices ($a=1,2,3$) at different  times $t = \{0.1, 0.5, 1.0, 1.5\}$~s. The simulations are initialized by assigning an identical Gaussian probability profile to all states:
\begin{equation}\label{eq:initial_gaussian_state}
P_a(x, 0) \sim \mathcal{N}(\mu_0 = 5, \sigma_0^2 = 1) \,, \quad \forall a \in \{1, 2, 3\} \,.
\end{equation}
We solve the system under two distinct drift-diffusion configurations (denoted as case (a) and case (b) in Fig.~\ref{fig:coupled_uncoupled_comparison}) to study also the differences arising from the parameter choice.

When the equations are decoupled ($g=0$), the system shows the classical, well-known behaviour of independent stochastic processes. Each probability curve acts as an isolated statistical entity: the local drift simply shifts the entire distribution toward the origin ($x=0$) over time, while the local diffusion causes a symmetric broadening and a gradual lowering of the peak. In this case, each state evolution remains entirely independent, and the profiles asymptotically decay to zero at the spatial boundaries.

Once the gauge coupling is activated at a representative strength ($g = 0.2$), a highly non-trivial dynamical behaviour emerges, showing a distinct and effective departure from the uncoupled reference. The most prominent qualitative effect is the severe suppression and flattening of the probability peaks. In the interacting model, the maximum amplitudes of the distributions are significantly squeezed downward compared to the independent case. This flattening is not caused by an increase in standard physical diffusion, but rather by a continuous, dynamical redistribution of probability among the states. The coupling forces the different sectors to constantly exchange probability flux, preventing any single distribution from maintaining a highly localized and sharp peak.

In addition to this flattening, the interaction introduces also a form of offset. This is visually evident in the profile of the $P_2(x)$ sector (green curve) in Fig. \ref{fig:coupled_uncoupled_comparison}(a), which undergoes a rigid vertical shift and establishes a constant, non-zero background across the entire spatial domain instead of decaying to zero at the boundaries. More generally, we note that rather than translating independently at rates strictly determined by their individual drift and diffusion parameters, the curves exert a mutual physical influence on one another. Furthermore, by directly comparing case (a) and case (b), we observe that inverting the diffusion coefficients changes the peak evolution of the $P_2(x)$ and $P_3(x)$ profiles.

\begin{figure}[t!]
    \centering
    % === IMMAGINE A SINISTRA (Case a) ===
    \begin{minipage}[t]{0.48\textwidth}
        \centering
        \includegraphics[width=\textwidth]{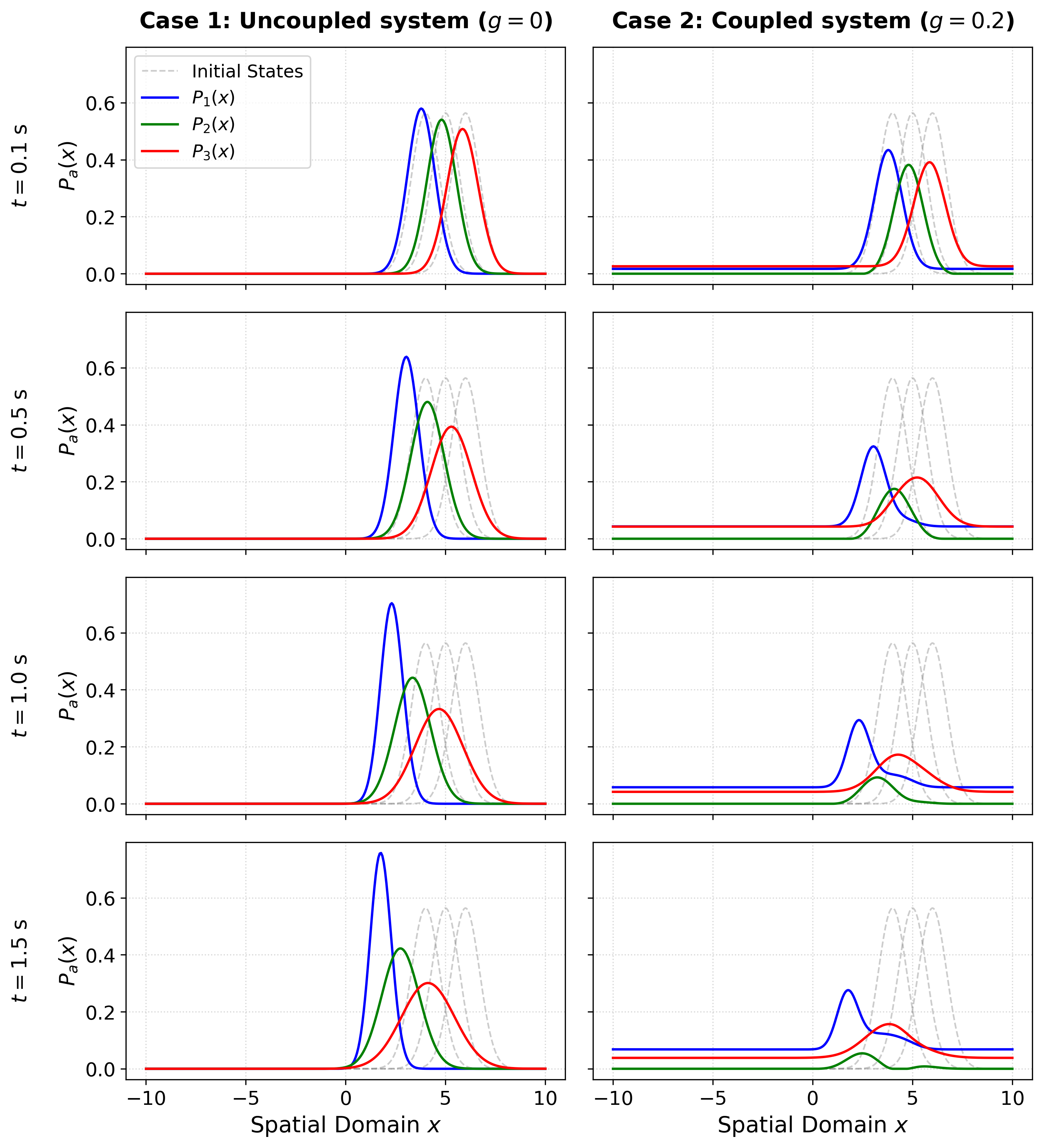}
        \vspace{1mm}
        \\ {(c)}
        \label{fig:case_3gauss}
    \end{minipage}%
    \hfill 
    % === IMMAGINE A DESTRA (Case b) ===
    \begin{minipage}[t]{0.48\textwidth}
        \centering
        \includegraphics[width=\textwidth]{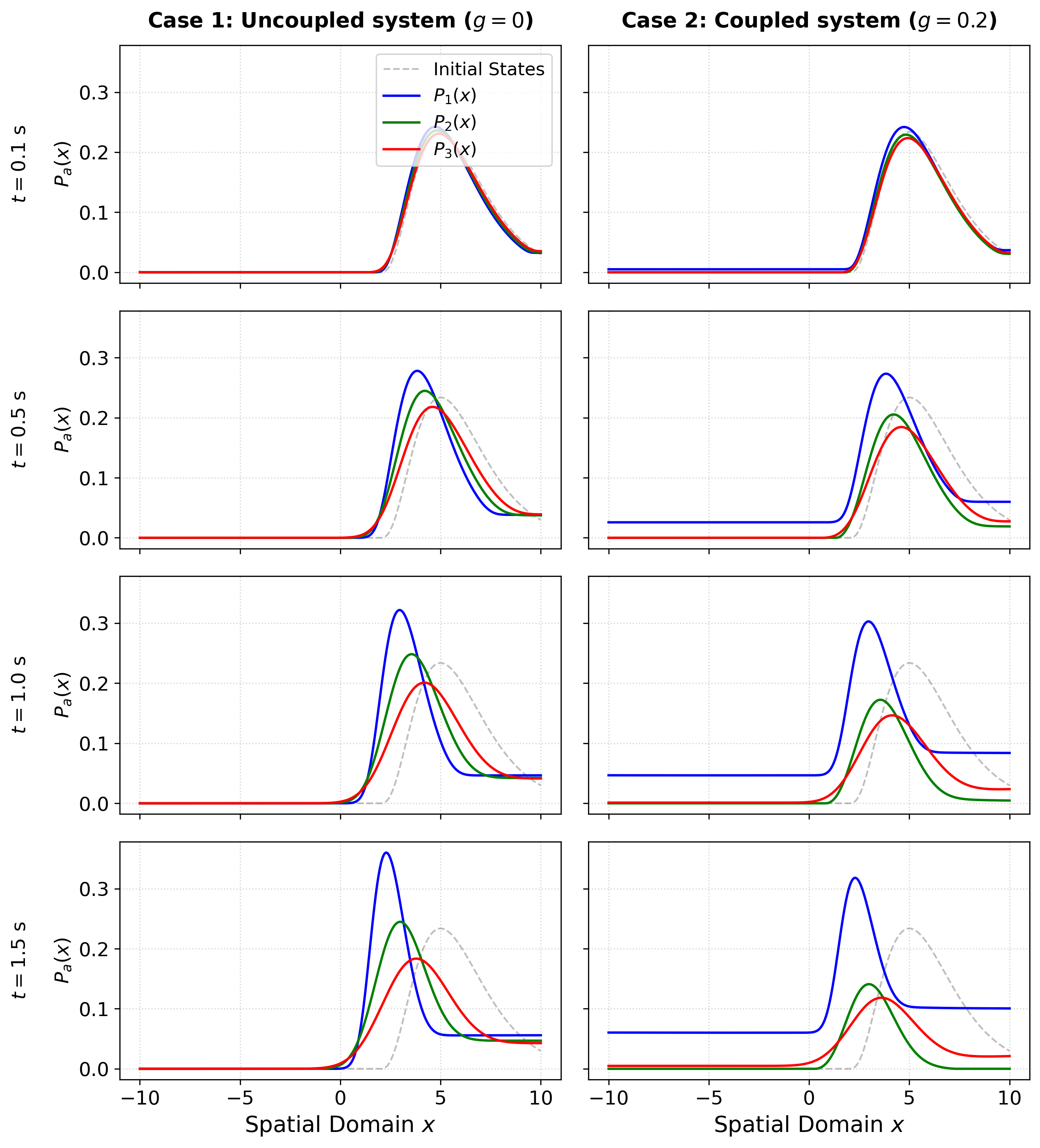}
        \vspace{1mm}
        \\ {(d)}
        \label{fig:case_gamma}
    \end{minipage}
    \caption{Temporal evolution of the probability density profiles $P_a(x)$ for the three sectors ($a=1,2,3$) (blue, green and red) under different initial conditions. Panel (c) shows the evolution starting from three spatially separated Gaussian profiles centered at $\mu_1 = 4.0$, $\mu_2 = 5.0$, and $\mu_3 = 6.0$. Panel (d) shows the evolution starting from an identical initial Gamma distribution profile, with shape parameter $\alpha = 4.0$ and rate parameter $\beta = 1.0$. In both panels, the left columns represent the uncoupled model ($g=0$) and the right columns illustrate the gauge-coupled model ($g=0.2$).}\label{fig:coupled_initial_conditions_sensitivity}
 \end{figure}

To further test the response of the proposed equations under different starting conditions, in Fig.~\ref{fig:coupled_initial_conditions_sensitivity} we consider two additional configurations where we break the symmetric initial setup. In the first case, denoted as panel (c), the system is initialized with three distinct Gaussian distributions centered at different positions, $P_1(x, 0) \sim \mathcal{N}(4, 1)$, $P_2(x, 0) \sim \mathcal{N}(5, 1) $, $P_3(x, 0) \sim \mathcal{N}(6, 1)$. In the second case, corresponding to panel (d), we initialize all three sectors using an identical, asymmetric Gamma probability density profile to study how the system evolves from a highly skewed initial state:
\begin{equation}\label{eq:gamma_initial_state}
P_a(x, 0) = \frac{\beta^\alpha}{\Gamma(\alpha)} x^{\alpha-1} e^{-\beta x} \,, \quad \forall a \in \{1, 2, 3\} \,,
\end{equation}
where $\alpha $ and $\beta$ represent the shape and rate parameters of the distribution, respectively, and $\Gamma(\alpha)$ is the Euler Gamma function. 
When the three initial peaks are separated in space, as shown in Fig.~\ref{fig:coupled_initial_conditions_sensitivity}(c), the uncoupled system ($g=0$) shows each curve shifting and broadening independently, preserving their spatial separation during the drift. However, when the coupling is turned on ($g=0.2$), this spatial separation is rapidly lost. The peaks of the green and red sectors are heavily squashed, transferring their probability to the blue sector, while a noticeable vertical offset develops on the background, preventing the curves from decaying to zero at the boundaries.

A similarly distinct behaviour appears for the initial Gamma profile in Fig.~\ref{fig:coupled_initial_conditions_sensitivity}(d). Without coupling, the initial asymmetry smoothly diffuses away as the independent curves drift toward the origin. When the interaction is activated, the coupled model completely redefines this evolution. The blue curve ($P_1$) experiences a strong localization, developing a very high and sharp peak that resists flattening over time. At the same time, the remaining sectors are suppressed, and a prominent vertical offset emerges at the domain boundaries, showing that the coupling can generate strong structural asymmetries depending on the initial profile.

\subsection{Asymptotic behaviour and constant coupling regimes }

To better characterize the properties of the gauge-coupled Fokker--Planck system, we extend our analysis to its asymptotic long-time behaviour and examine how the dynamics scale with the coupling strength $g$. 

\begin{figure}[t!]
    \centering
    \includegraphics[width=0.6\textwidth]{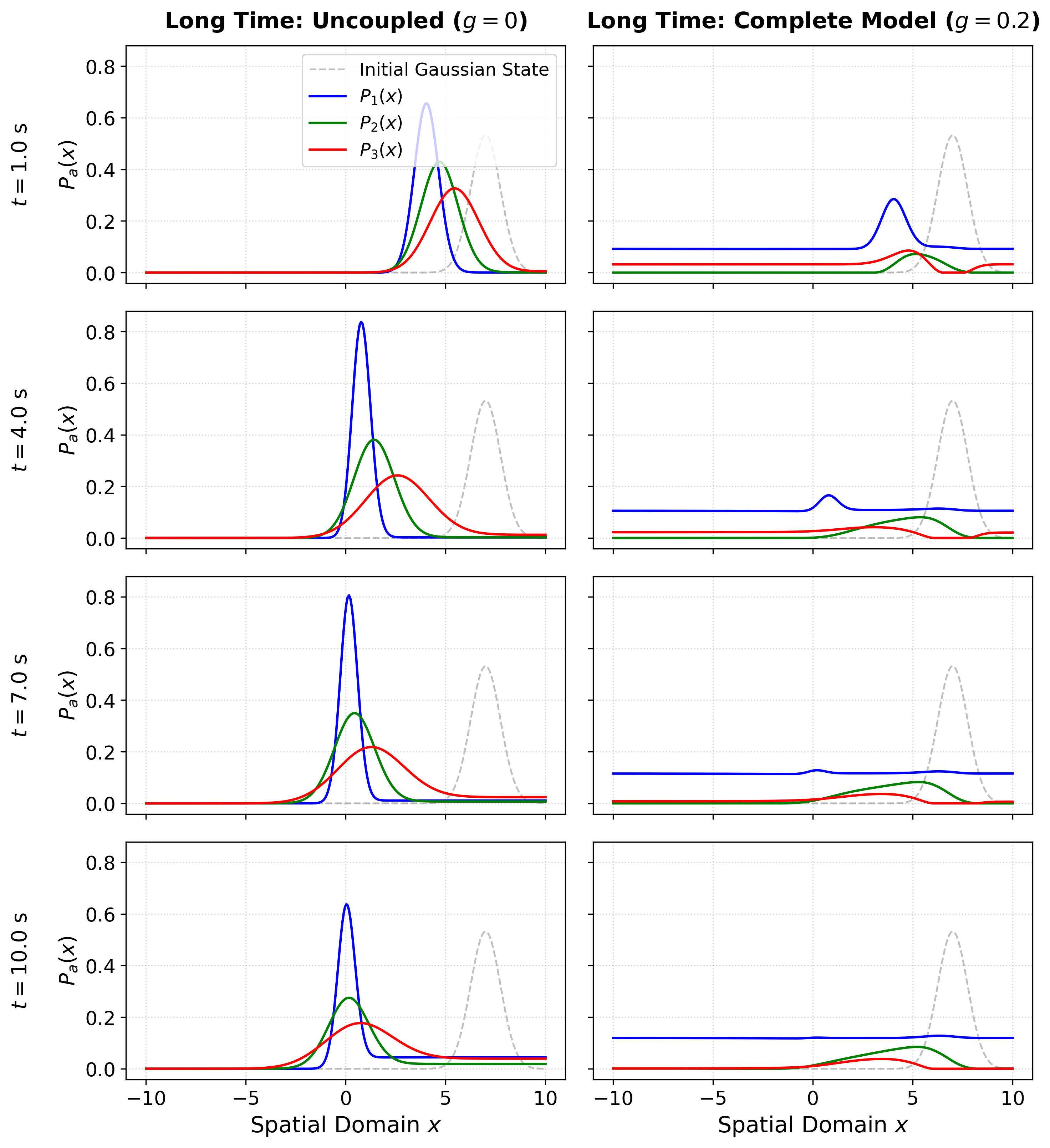}
    \vspace{2mm}
    \caption{Long-time asymptotic relaxation towards the origin ($x=0$) comparing the uncoupled linear model (solid lines, $g=0$) against the complete quadratic model (dashed lines, $g=0.2$). The single-column subplots track sequential temporal snapshots up to the asymptotic regime at physical times $t = 1.0, 4.0, 7.0, 10.0$~s. All three sectors ($a=1,2,3$) initiate from an identical overlapping Gaussian profile centered at $x = 7.0$ (gray dashed line) under fixed drift rates $\gamma_a = (0.55, 0.40, 0.25)$ and diffusion strengths $D = (0.1, 0.4, 0.7)$.}
    \label{fig:long_time_gaussian_overlay}
\end{figure}

We first investigate the long-time relaxation dynamics of the system. Figure~\ref{fig:long_time_gaussian_overlay} shows the evolution of the probability profiles from their initial state toward the asymptotic limit. 
In the uncoupled case, the three sectors exhibit standard independent relaxation: under the influence of the linear drift, the peaks slide toward the origin ($x=0$), while diffusion induces symmetric broadening. By $t = 10.0$~s, the uncoupled system has effectively reached its localized stationary state, where drift and diffusion balance out, and the profiles decay to zero at the boundaries. 

In contrast, the complete coupled model ($g=0.2$) exhibits a completely different asymptotic behavoiur. The coupling prevents full localized relaxation at the origin. Instead of decaying at the boundaries, all three probability densities develop a flat-like distribution. At $t = 10.0$~s, the system reaches a steady-state configuration characterized by a more flat distribution, with the blue curve $P_1$ stabilizing at the highest value, followed by $P_2$ in green and $P_3$ in red. This confirms that the gauge interaction structurally reorganizes the asymptotic configuration of the system, forcing a transition from local confinement to global delocalization in the state space.

\begin{figure}[b!]
    \centering
    \includegraphics[width=0.8\textwidth]{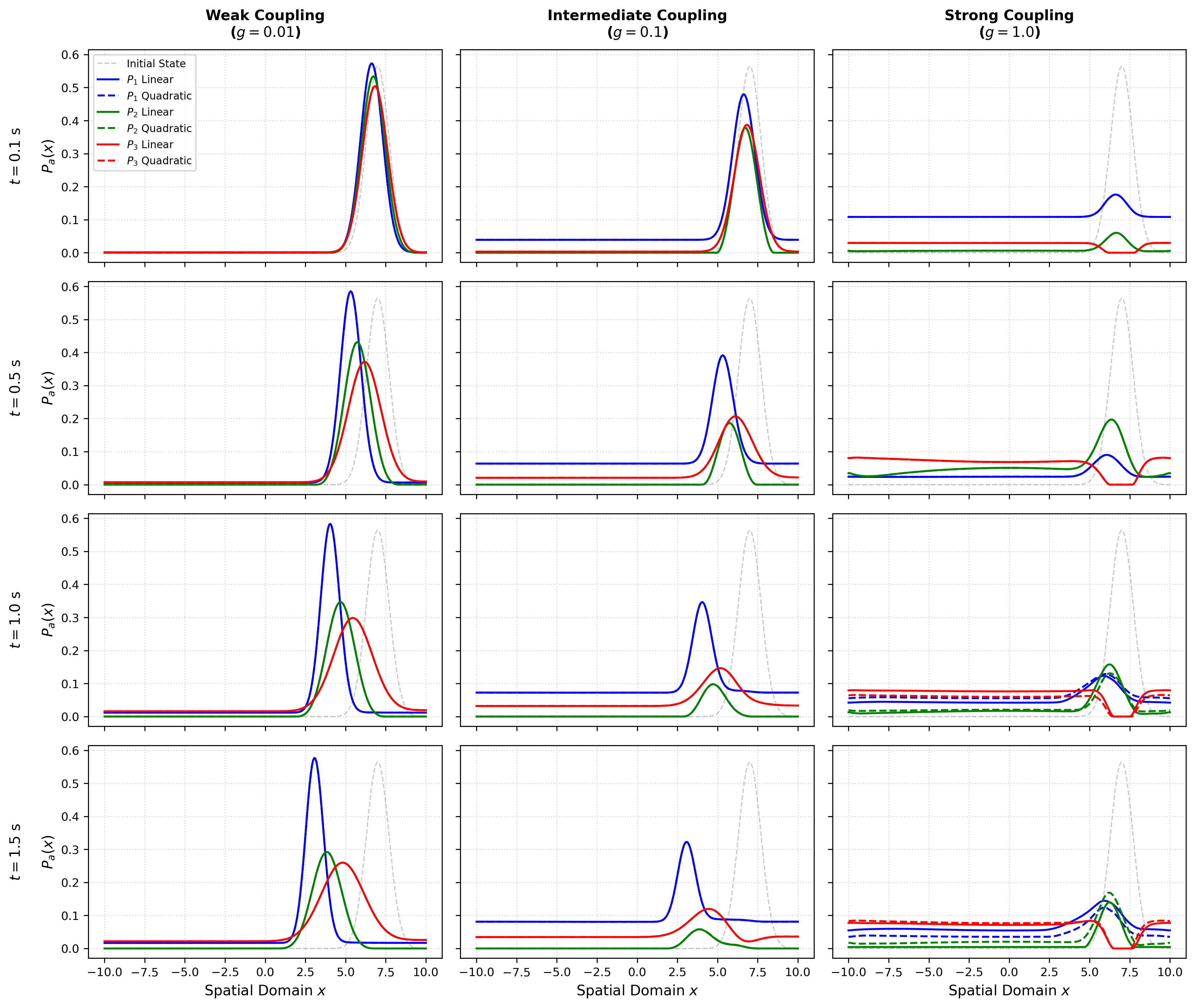}
    \vspace{2mm}
    \caption{Temporal evolution of the probability density profiles $P_a(x)$ for the three sectors ($a=1,2,3$) comparing the truncated linear coupling model (solid lines, $g^2 = 0$) against the complete quadratic model (dashed lines, including $g^2$ contributions). The matrix columns illustrate three distinct coupling strengths: weak coupling ($g = 0.01$), intermediate coupling ($g = 0.1$), and strong coupling ($g = 1.0$), while rows track snapshots from top to bottom at physical times $t = 0.1, 0.5, 1.0, 1.5$~s. All simulations start from an identical initial Gaussian distribution centered at $x = 7.0$ (gray dashed line) with fixed drift coefficients $\gamma_a = (0.55, 0.40, 0.25)$ and diffusion coefficients $D = (0.1, 0.4, 0.7)$.}
    \label{fig:linear_vs_quadratic_overlay}
\end{figure}

A second key aspect, we investigate is how the intensity of costant coupling $g$ change the system evolution. In Fig.~\ref{fig:linear_vs_quadratic_overlay}, we systematically vary the coupling strength from weak ($g = 0.01$) to intermediate ($g = 0.1$) and strong ($g = 1.0$) regimes. At the same time, we compare the truncated linear coupling model ($g^2 = 0$) against the complete quadratic model (including $g^2$ terms) to verify under which conditions the $g^2$ terms can be safely neglected to simplify the governing equations.

Under weak coupling, the system behaves very similarly to the uncoupled case. The interaction is too weak to alter the independent transport, and the profiles drift and diffuse toward the left with negligible exchange of probability density between the states. At the intermediate coupling level, we observe the onset of the coupling effect: the local peaks begin to dampen and exhibit a reduced drift velocity compared to the weak coupling case, while a uniform background plateau starts to lift the distributions of the blue ($P_1$) and red ($P_3$) curves from the zero-boundary condition. 

When the system enters the strong coupling regime ($g = 1.0$, right column of Fig.~\ref{fig:linear_vs_quadratic_overlay}), the dynamics are almost entirely governed by the coupling source. The local peaks are drastically suppressed and flattened within the very first moments of the evolution ($t = 0.1$~s). The probability is rapidly homogenised across the domain, establishing a flat distribution for all states.

Furthermore, this strong coupling column highlights the critical role of the quadratic $g^2$ contributions in our numerical scheme. In the weak ($g=0.01$) and intermediate ($g=0.1$) regimes, the linear and quadratic models perfectly overlap. This overlap clearly indicates that for small to moderate coupling strengths, the $g^2$ term is indeed negligible, allowing for a simplified linear formulation of the equations. Conversely, in the strong coupling case ($g=1.0$), the linear model (solid lines) and the complete quadratic model (dashed lines) visibly diverge, showing significant differences. This sensitive divergence at high $g$ values confirms that the quadratic terms are essential to preserve the consistency and physical behaviour of the model when the gauge interaction dominates.

However, the numerical implementation faces strict stability boundaries when the coupling is pushed further. For very strong coupling strengths ($g \gg 1$), the numerical system becomes highly unstable. In this extreme regime, the non-local source terms in the operator splitting scheme grow excessively large relative to the physical drift and diffusion. This leads to the rapid development of high-frequency, non-physical spatial oscillations across the flat distributions of all states, eventually causing the simulation to diverge. This behaviour underscores the necessity of maintaining a balanced coupling regime to preserve the stability and physical validity of the numerical scheme.

\section{Conclusions}\label{Conclusions}
In this work, we presented the explicit numerical implementation of a non-Abelian $SU(2)$ gauge-coupled Fokker--Planck framework designed for multi-state stochastic dynamics. By projecting the formulation into a controlled $(1+1)$-dimensional space, we transformed the abstract integro-differential system into a robust computational tool. Coupling an explicit operator-splitting scheme with the non-oscillatory MPDATA advection solver we ensure strict positivity preservation and total probability conservation, while avoiding numerical oscillations, displaying predictable first-order temporal-spatial error convergence.

The non-Abelian interaction induces dynamic cross-talk between distinct probability fields through non-local spatio-temporal terms. In contrast to independent Fokker--Planck evolution, the coupled dynamics lead to peak suppression, spatial broadening, and structural offsets across state distributions. In the asymptotic long-time limit, the system undergoes a fundamental transition from localized stationary states centred at the potential minima to  spatial delocalization within the finite computational domain across the coupled states. 
Systematic scaling of the coupling constant $g$ reveals distinct operational regimes. For weak to intermediate coupling ($g\lesssim 0.1$), the linear model ($g^2 \rightarrow0$) accurately captures the dynamics. However, strong coupling regimes ($g\simeq1.0$) necessitate the full quadratic formulation to preserve physical consistency, up to stability limits where non-local source terms dominate advective-diffusive transport. 

Finally, the algorithmic architecture established in this work aims to bridge a critical computational gap, providing a scalable path toward incorporating multi-habit ice crystal size evolutions into cloud microphysical models and Global Climate Models (GCMs). Beyond atmospheric cloud microphysics and radiative transfer applications, the unified mathematical structure of gauge-coupled Fokker--Planck equations offers a versatile paradigm for describing multi-state stochastic evolution in complex physical, biological, and financial processes.

Looking forward, future work will pursue several promising directions. On the cloud physics front, the framework will be extended to higher spatial dimensions ($D=2$) to construct realistic phase-space representations of ice crystal habit evolution in cirrus clouds. This step will involve directly coupling the internal symmetry algebra with macroscopic thermodynamic variables and validating model outputs against observational microphysical datasets to rigorously assess its performance for climate applications. At the same time, it will be of great interest to explore the general applicability of these gauge-coupled equations across other disciplines, systematically investigating how this framework can model complex transitions in neural dynamics, financial processes, and other domains where stochastic dynamics and Fokker--Planck formulations are widely utilised, ultimately uncovering the broader physical consequences of non-Abelian interactions in stochastic systems and how they can enhance predictive modelling capabilities.

\subsection*{Acknowledgments}
\small This work is supported by the Spanish Government and ERDF/EU-Agencia Estatal de Investigaci\'on MCIN/AEI/10.13039/501100011033, Grants No.~PID2023-146220NB-I00, No.~EUR2025-164820, and No.~CEX2023-001292-S. LC is supported by Generalitat Valenciana plan GenT program (CIDEGENT/2020/011). Part of the numerical calculations were carried out on the Funes computing cluster, hosted at the computational infrastructure of the Instituto de F\'isica Corpuscular (IFIC). This work is also supported by the CASIA project (CUP: F93C23000430001), funded by the Italian Space Agency (ASI) .

\subsection*{Code Availability}
\small The numerical framework and the source code used to generate the simulations presented in this study are openly available on Zenodo at \url{https://zenodo.org/records/22956953}.
\newpage

\appendix

\section{Explicit 1D Coupled System for Numerical Implementation}\label{Appendix_Integrals}

This Appendix provides the complete algebraic expansion and serves as an external, self-contained reference for all the calculus and mathematical derivations underlying the $(1+1)$-dimensional reduction under $SU(2)$ symmetry. The expressions detailed below unfold every coupled term explicitly, acting as the direct blueprint for the discretization routines implemented in our numerical framework.

For a generic probability density component $P_a$ within an $SU(2)$ gauge framework, the projection into a single spatial coordinate $x$ leads to the following integro-differential master equation:
\begin{equation}\label{App_general_1D}
\begin{aligned}
\partial_t P_a &= -\partial_x J_a \\
&+ g \ \epsilon_{abc} \left[ \int \mathrm{d}x \, \big( -P_c \partial_x P_b  - (\partial_t P_b) J_c +  2 P_c (\partial_t J_b)\big) + \int \mathrm{d}t \, \big( -J_c \partial_t J_b - (\partial_x J_b) P_c + 2 J_c (\partial_x P_b) \big) \right] \\
& + g^2  \left[ \int \mathrm{d}x \, P_b ( P_b J_a - P_a J_b ) + \int \mathrm{d}t \, J_b ( J_b P_a - J_aP_b) \right] \,.
\end{aligned}
\end{equation}

Evaluating Eq.~\eqref{App_general_1D} across the full set of generators ($a \in \{1,2,3\}$) yields a closed system of three coupled partial integro-differential equations. Retaining the compact current notation $J_a(x,t)$, the system expands into the following structural form:
\begin{equation}\label{App_SU2_current_system}
\left\{
\begin{aligned}
\partial_t P_1 &= -\partial_x J_1 \\
&+ g \left[ \int \mathrm{d}x \, \big( - P_3 \partial_x P_2   - (\partial_t P_2) J_3 + 2 P_3 \partial_t J_2\big) + \int \mathrm{d}t \, \big( - J_3 \partial_t J_2 - (\partial_x J_2) P_3 + 2 J_3 (\partial_x P_2) \big) \right] \\
& - (2 \leftrightarrow 3) + g^2 \left[ \int \mathrm{d}x \, P_2 (  P_2 J_1 -P_1 J_2 ) + \int \mathrm{d}t \, J_2 (  J_2 P_1 -   J_1 P_2 ) \right] + (2 \leftrightarrow 3) \,,\\[12pt]
\partial_t P_2 &= -\partial_x J_2 \\
&+ g \left[ \int \mathrm{d}x \, \big( -P_1 \partial_x P_3   - (\partial_t P_3) J_1 + 2 P_1 \partial_t J_3\big) + \int \mathrm{d}t \, \big(- J_1 \partial_t J_3 - (\partial_x J_3) P_1 + 2 J_1 (\partial_x P_3) \big) \right] \\
&- (3 \leftrightarrow 1) + g^2 \left[ \int \mathrm{d}x \, P_3 ( P_3 J_2 - P_2 J_3) + \int \mathrm{d}t \, J_3 (  J_3 P_2  -  J_2 P_3) \right] + (3 \leftrightarrow 1) \,,\\[12pt]
\partial_t P_3 &= -\partial_x J_3 \\
&+ g \left[ \int \mathrm{d}x \, \big( -P_2 \partial_x P_1  - (\partial_t P_1) J_2 + 2 P_2 \partial_t J_1 \big) + \int \mathrm{d}t \, \big( -J_2 \partial_t J_1 - (\partial_x J_1) P_2 + 2 J_2 (\partial_x P_1) \big) \right] \\
&- (1 \leftrightarrow 2) + g^2 \left[ \int \mathrm{d}x \, P_1 (  P_1 J_3 -P_3 J_1 ) + \int \mathrm{d}t \, J_1 (   J_1 P_3 - J_3 P_1 ) \right] + (1 \leftrightarrow 2) \,.
\end{aligned}
\right.
\end{equation}

To provide the explicit mathematical reference required for quantitative modelling and algorithmic tracking, we substitute the definition of the local, one-dimensional spatial currents—given by $J_a(x, t) = D_{(a)}^x(x) P_a(x, t) - \mathbb{D}_{(a)}^{xx} \partial_x P_a(x, t)$—directly into Eq.~\eqref{App_SU2_current_system}. This fully unrolls all non-local drift and cross-diffusion dependencies, resulting in the following uncompressed, definitive system of equations:
\begin{equation}\label{App_SU2_fully_expanded}
\left\{
\begin{aligned}
\partial_t P_1 &= -\partial_x \bigl( D_{(1)}^x P_1 \bigr) + \mathbb{D}_{(1)}^{xx} \partial_x^2 P_1  \\
&+ g  \int \mathrm{d}x \Big[- P_3 \partial_x P_2  +2 P_3 \partial_t \bigl( D^x_{(2)}P_2 - \mathbb{D}_{(2)}^{xx} \partial_x P_2 \bigr) - (\partial_t P_2) \bigl( D^x_{(3)}P_3 - \mathbb{D}_{(3)}^{xx} \partial_x P_3 \bigr) \Big] \\
&+ g  \int \mathrm{d}t \Big[- \bigl( D^x_{(3)}P_3 - \mathbb{D}_{(3)}^{xx} \partial_x P_3 \bigr) \partial_t \bigl( D^x_{(2)}P_2 - \mathbb{D}_{(2)}^{xx} \partial_x P_2 \bigr) - \partial_x \bigl( D^x_{(2)}P_2 - \mathbb{D}_{(2)}^{xx} \partial_x P_2 \bigr) P_3 \\
&\qquad \qquad \ + 2 \bigl( D^x_{(3)}P_3 -\mathbb{D}_{(3)}^{xx} \partial_x P_3 \bigr) (\partial_x P_2)  \Big] - \Big(2 \leftrightarrow 3\Big)\\
& + g^2 \int \mathrm{d}x \, \Big [ - P_2 \big( P_1 (D^x_{(2)}P_2 - \mathbb{D}^{xx}_{(2)}\partial_x P_2) - P_2 (D^{x}_{(1)}P_1 - \mathbb{D}^{xx}_{(1)}\partial_x P_1) \big) \Big ] \\
& + g^2 \int \mathrm{d}t \, \Big [ - (D^{x}_{(2)}P_2 - \mathbb{D}^{xx}_{(2)}\partial_x P_2) \big( P_2 (D^{x}_{(1)}P_1 - \mathbb{D}^{xx}_{(1)}\partial_x P_1) + P_1 (D^{x}_{(2)}P_2 - \mathbb{D}^{xx}_{(2)}\partial_x P_2) \big) \Big ]\\
&+ \Big(2 \leftrightarrow 3\Big) \\[15pt]
\partial_t P_2 &= -\partial_x \bigl( D_{(2)}^x P_2 \bigr) + \mathbb{D}_{(2)}^{xx} \partial_x^2 P_2  \\
&+ g  \int \mathrm{d}x  \Big[- P_1 \partial_x P_3  +2 P_1 \partial_t \bigl( D^x_{(3)}P_3 - \mathbb{D}_{(3)}^{xx} \partial_x P_3 \bigr) - (\partial_t P_3) \bigl( D^x_{(1)}P_1 - \mathbb{D}_{(1)}^{xx} \partial_x P_1 \bigr) \Big]  \\
&+ g  \int \mathrm{d}t \Big[ -\bigl( D^x_{(1)}P_1 - \mathbb{D}_{(1)}^{xx} \partial_x P_1 \bigr) \partial_t \bigl( D^x_{(3)}P_3 - \mathbb{D}_{(3)}^{xx} \partial_x P_3 \bigr) - \partial_x \bigl( D^x_{(3)}P_3 - \mathbb{D}_{(3)}^{xx} \partial_x P_3 \bigr) P_1 \\
&\qquad \qquad \ + 2 \bigl( D^x_{(1)}P_1 - \mathbb{D}_{(1)}^{xx} \partial_x P_1 \bigr) (\partial_x P_3)  \Big] - \Big (3 \leftrightarrow 1\Big) \\
& + g^2 \int \mathrm{d}x \, \Big [ - P_3 \big( P_2 (D^{x}_{(3)}P_3 - \mathbb{D}^{xx}_{(3)}\partial_x P_3) - P_3 (D^{x}_{(2)}P_2 - \mathbb{D}^{xx}_{(2)}\partial_x P_2) \big) \Big ] \\
& + g^2 \int \mathrm{d}t \, \Big [ - (D^{x}_{(3)}P_3 - \mathbb{D}^{xx}_{(3)}\partial_x P_3) \big( P_3 (D^{x}_{(2)}P_2 - \mathbb{D}^{xx}_{(2)}\partial_x P_2) - P_2 (D^{x}_{(3)}P_3 - \mathbb{D}^{xx}_{(3)}\partial_x P_3) \big) \Big ]\\
&+\Big (3 \leftrightarrow 1\Big)  \\[15pt]
\partial_t P_3 &= -\partial_x \bigl( D_{(3)}^x P_3 \bigr) + \mathbb{D}_{(3)}^{xx} \partial_x^2 P_3  \\
&+ g  \int \mathrm{d}x  \Big[ - P_2 \partial_x P_1  +2 P_2 \partial_t \bigl( D^x_{(1)}P_1 - \mathbb{D}_{(1)}^{xx} \partial_x P_1 \bigr) - (\partial_t P_1) \bigl( D^x_{(2)}P_2 - \mathbb{D}_{(2)}^{xx} \partial_x P_2 \bigr) \Big] \\
&+ g  \int \mathrm{d}t \Big[ -\bigl( D^x_{(2)}P_2 - \mathbb{D}_{(2)}^{xx} \partial_x P_2 \bigr) \partial_t \bigl( D^x_{(1)}P_1 - \mathbb{D}_{(1)}^{xx} \partial_x P_1 \bigr) - \partial_x \bigl( D^x_{(1)}P_1 - \mathbb{D}_{(1)}^{xx} \partial_x P_1 \bigr) P_2 \\
&\qquad \qquad \ + 2 \bigl( D^x_{(2)}P_2 - \mathbb{D}_{(2)}^{xx} \partial_x P_2 \bigr) (\partial_x P_1)  \Big] - \Big(1 \leftrightarrow 2\Big) \\
& + g^2 \int \mathrm{d}x \, \Big [ - P_1 \big( P_3 (D^{x}_{(1)}P_1 - \mathbb{D}^{xx}_{(1)}\partial_x P_1) - P_1 (D^{x}_{(3)}P_3 - \mathbb{D}^{xx}_{(3)}\partial_x P_3) \big) \Big ] \\
& + g^2 \int \mathrm{d}t \, \Big [ - (D^{x}_{(1)}P_1 - \mathbb{D}^{xx}_{(1)}\partial_x P_1) \big( P_1 (D^{x}_{(3)}P_3 - \mathbb{D}^{xx}_{(3)}\partial_x P_3) - P_3 (D^{x}_{(1)}P_1 - \mathbb{D}^{xx}_{(1)}\partial_x P_1) \big) \Big ]\\
& \Big(1 \leftrightarrow 2\Big) 
\end{aligned}
\right.
\end{equation}

\newpage

\bibliographystyle{unsrt}
\bibliography{references.bib}

\end{document}